# Controllable retinal image synthesis using conditional StyleGAN and latent space manipulation for improved diagnosis and grading of diabetic retinopathy

Somayeh Pakdelmoez[a], Saba Omidikia[a], Seyyed Ali Seyyedsalehi[a,*], Seyyede Zohreh Seyyedsalehi[b]

*[a]Department of Biomedical Engineering, Amirkabir University of Technology, Tehran, Iran*
*[b]Department of Biomedical Engineering, Faculty of Health, Tehran Medical Sciences, Islamic Azad University, Tehran, Iran*

## Abstract

Diabetic retinopathy (DR) is a consequence of diabetes mellitus characterized by vascular damage within the retinal tissue. Timely detection is paramount to mitigate the risk of vision loss. However, training robust grading models is hindered by a shortage of annotated data, particularly for severe cases. This paper proposes a framework for controllably generating high-fidelity and diverse DR fundus images, thereby improving classifier performance in DR grading and detection. We achieve comprehensive control over DR severity and visual features (optic disc, vessel structure, lesion areas) within generated images solely through a conditional StyleGAN, eliminating the need for feature masks or auxiliary networks. Specifically, leveraging the SeFa algorithm to identify meaningful semantics within the latent space, we manipulate the DR images generated conditionally on grades, further enhancing the dataset diversity. Additionally, we propose a novel, effective SeFa-based data augmentation strategy, helping the classifier focus on discriminative regions while ignoring redundant features. Using this approach, a ResNet50 model trained for DR detection achieves 98.09% accuracy, 99.44% specificity, 99.45% precision, and an F1-score of 98.09%. Moreover, incorporating synthetic images generated by conditional StyleGAN into ResNet50 training for DR grading yields 83.33% accuracy, a quadratic kappa score of 87.64%, 95.67% specificity, and 72.24% precision. Extensive experiments conducted on the APTOS 2019 dataset demonstrate the exceptional realism of the generated images and the superior performance of our classifier compared to recent studies.

## 1. Introduction

Diabetic retinopathy (DR) is a complication caused by diabetes that damages the blood vessels in the light-sensitive tissue of the retina due to high blood sugar levels. The diagnosis of DR primarily relies on the assessment of retinal lesions, including microaneurysms, hemorrhages, neovascularization, and hard exudates, visualized through fundus photography [1]. Ophthalmologists typically identify and grade the severity of DR by examining the type and quantity of associated lesions. The international protocol categorizes the severity of DR into five levels: normal, mild, moderate, severe non-proliferative DR (NPDR), and proliferative DR (PDR) [2]. Delayed diagnosis of lower-grade non-proliferative cases can exacerbate the condition, escalating the risk of developing PDR, the most severe form of the disease. In this stage, abnormal blood vessels proliferate on the retinal surface, leading to bleeding, scar tissue formation, and potential retinal detachment. If left untreated, this condition can cause significant irreversible visual impairment and permanent blindness [3].

---
*Corresponding author
 Email address:* `ssalehi@aut.ac.ir` (S.A. Seyyedsalehi)

Wykof et al. [4] studied more than 53,000 patients with newly diagnosed DR with good vision at diagnosis for the development of irreversible blindness. Their analysis revealed that eyes with moderate NPDR, severe NPDR, and PDR at the time of diagnosis were 2.6, 3.6, and 4.0 times more likely, respectively, to develop sustained blindness after 2 years compared to eyes with mild DR at diagnosis. This highlights the paramount importance of accurate and prompt diagnosis of each abnormal grade. However, the process of diagnosing and grading DR is time-consuming and arduous for ophthalmologists.

Computer-aided diagnosis (CAD) systems utilizing deep learning algorithms have demonstrated promising results in analyzing color fundus photographs, assisting ophthalmologists in reducing workload and enhancing diagnostic accuracy for DR [5], [6], [7]. Nevertheless, developing a robust DR diagnosis model with high accuracy requires a substantial amount of diverse and well-balanced data. Unfortunately, publicly accessible datasets often exhibit imbalanced DR class distribution due to privacy concerns limiting the availability of abnormal cases, which are inherently less frequent compared to normal ones [8]. Training models on such imbalanced datasets leads to decreased sensitivity towards abnormal samples, hindering their ability to detect these crucial cases. However, accurately and promptly diagnosing abnormalities associated with each grade is crucial to mitigate negative consequences. This study addresses this challenge by employing effective data augmentation techniques to synthesize DR images for all grades in a controllable process. Our approach allows for manipulation of both general retinal image features and those specific to DR pathology. Through participation in the detection and grading task, we demonstrate the practicality and efficacy of our method in enhancing diagnostic performance.

Generative adversarial network (GAN) [9], a prominent class of probabilistic generative models, has demonstrated significant potential in various image synthesis tasks. These include text-to-image generation [10], image editing [11], image-to-image translation [12], and video generation [13]. GANs achieve this by learning the underlying data distribution and generating realistic images from random noise. However, vanilla GANs lack control over the generated content. To address this limitation, conditional GANs (cGANs) [14] have been developed, incorporating additional information to guide and control the image generation process. Pix2Pix [15] is a prominent example, where the conditional input is an image, enabling powerful image-to-image translation tasks.

Several studies have utilized cGAN and Pix2Pix to synthesize retinal images of good quality, employing a controllable process for image generation [16], [17], [18]. However, these approaches often focus on controlling only specific features within the generated image. Additionally, they typically require pre-existing segmentation masks corresponding to the desired controlled features. Furthermore, images corresponding to the masks are required for the training process. In some studies, masks are synthesized from noise using another GAN model. In others, a segmentation model specific to the desired feature (e.g., vessels, optic disc, lesion areas) needs to be trained beforehand. This dependence on masks for controllable synthesis presents several challenges. First, it limits the diversity of generated images due to the reliance on the availability of diverse masks. Second, the need to pre-train a separate system (mask generation or segmentation) increases the overall complexity of the process. Finally, the quality of the generated image is inherently linked to the performance of the mask generation or segmentation system.

One of the recent advancements in GANs design is StyleGAN [19], inspired by style transfer techniques. This approach enhances image generation quality through the incorporation of progressive training and adaptive instance normalization. The StyleGAN architecture grants control over both high-level semantic attributes and fine-grained details within the generated images. This unique architectural design enables improved feature separation within a more disentangled latent space, resulting in greater semantic interpretability compared to conventional GANs. Consequently, this capability facilitates more effective image manipulations based on the latent space, allowing for more extensive control over synthesized image features.

This paper introduces a novel approach that utilizes the conditional StyleGAN network to generate retinal images by controlling over DR grades. To enhance control over the various features of each class's generated images, we exclusively employ a pre-trained conditional generator, eliminating the need for pre-existing masks or auxiliary system pre-training. Concretely, we utilize the SeFa algorithm [20] to explore the rich latent space of the network unsupervisedly, identifying meaningful directions. This enables manipulation of significant features such as vessel structure, optic disc, and lesions, facilitating the generation of high-quality and diverse retinal images. We evaluate the effectiveness of the generated images both qualitatively and quantitatively through a wide range of comparative experiments and various metrics in both DR screening and grading tasks. Furthermore, we present a novel approach, referred to as effective SeFa-based data augmentation. This framework allows for effectively augmenting images through SeFa-based manipulation of generated images, significantly enhancing DR detection performance. The results confirm the high quality and diversity of the generated images by demonstrating their effectiveness in boosting classifier performance for both DR screening and grading when incorporated with existing real images. Additionally, the results demonstrate the superiority of our model compared to recent studies on both tasks.

This paper makes the following key contributions:

- We propose a novel approach to have comprehensive control in generating DR fundus images by modifying the state-of-the-art StyleGAN into a conditional structure. This allows for controllable synthesis of DR images with different grades using the APTOS 2019 dataset. To achieve further control over other semantic features, we manipulate the generated images through the SeFa approach, which identifies semantics encoded in the latent space. This method enables extensive control in DR image synthesis without requiring feature masks or auxiliary networks.

- Comparative experiments were conducted for vasculature and lesion manipulation based on the semantics discovered by SeFa and Style mixing manipulation. Qualitative assessments demonstrate the high quality and realistic nature of the synthesized images.

- Quantitative assessments across various metrics, incorporating the generated images in training a ResNet50 model, demonstrate the effectiveness of the synthesized images in both DR grading and diagnosis tasks, as well as their superiority over other recent studies.

- Specifically, for the grading of DR, expanding existing real images with the generated images boost accuracy, quadratic kappa score, sensitivity, specificity, precision, F1-score, and AUC-ROC by 7%, 10.61%, 18.9%, 1.7%, 16.27%, 20.92%, and 3.68%, respectively.

- We present a novel, effective SeFa-based data augmentation technique that significantly enhances the classifier's accuracy, specificity, precision and F1-score in DR detection to 98.09%, 99.44%, 99.45%, and 98.09%, respectively.

## 2. Related works

### 2.1. Generative Adversarial Networks

GANs have emerged as highly successful models for image generation [20]. GANs operate as unsupervised generative models comprising a generator and a discriminator, which engage in an adversarial learning process. The generator takes random noise as input, with the goal of generating realistic data that resembles the training set. The objective of the discriminator is to accurately distinguish between real and fake images, while the generator tries to deceive the discriminator by generating more realistic data. This competitive dynamic, resembling a min-max game, drives both networks to iteratively improve, ultimately resulting in the generation of higher-quality images [9]. A significant development in the realm of GANs is the introduction of conditional Generative Adversarial Networks (cGANs) [14]. Unlike vanilla GANs, which lack control over the state of generated data, cGAN leverages training data labels to generate images corresponding to specific class labels.

Moreover, modern GAN architectures frequently incorporate progressive growing into their training process to facilitate the generation of high-resolution images [19], [21]. Through further development of this progressive structure, StyleGAN has made significant advancements in generating near-photorealistic images of the human face, owing to its style-based generator architecture [19]. In this architecture, the generator no longer operates as a black box; instead, it gains the ability to interpret various aspects of the image synthesis process, enabling relative control over the features of the generated images. The intermediate latent space of StyleGAN facilitates better disentanglement of latent factors of variation. By employing the Adaptive Instance Normalization (AdaIN) operation across different layers of the generator, StyleGAN achieves enhanced control over the style and appearance of the generated images. Furthermore, the disentanglement of image content and style information enables style mixing, allowing for the combination of different styles from different images. This results in more diverse and controllable image synthesis. Additionally, the injection of controllable noise, with adjustable levels and distributions across different layers, grants control over stochasticity, thereby adding variation and fine-grained details to the generated images. Consequently, this leads to outputs that are more realistic and visually appealing. By leveraging a more disentangled latent space, StyleGAN achieves semantic interpretability. This characteristic enhances the power and effectiveness of latent space-based image manipulation, surpassing other forms of GANs in terms of controllability of image generation and visual quality.

Utilizing the controller structure present in both cGAN and StyleGAN networks, this paper, investigates a conditional extension to the StyleGAN architecture for the synthesis of DR images. This extension grants us comprehensive control over various aspects, including DR grade, general retinal image features, and DR-related features.

## 2.2. Control over vasculature in GAN-based retinal image generation

Many researchers have performed the synthesis of retinal images by controlling the structure of the vessels. In these studies, a Pix2Pix-based structure, a type of cGAN, is generally used to control the vessel structures of the synthesized images by translating the vessel mask into the fundus image. However, to ensure control over the vessel structures of the synthesized image, it is necessary to have pre-existing vessel masks. Moreover, in order to perform controllable synthesis with the minimum error, training the Image-to-Image translation models necessitates a significant number of image and mask pairs, while such extensive datasets are not publicly available. To address this issue, Costa et al. [22] and Chen et al [16] utilized a U-Net network [23] to segment images and generate masks, thus creating image and mask pairs. However, this additional system training for controllable synthesis introduces complexity. Furthermore, the quality of the synthesized images heavily relies on the performance of the segmentation network, thus a weak network can result in incomplete masks. Consequently, not only is accurate control over the vessel structure not achieved, but also the presence of spurious artifacts, chromatic noise, and occasionally incomplete field of view can be observed, leading to the generation of unrealistic low-quality images.

To mitigate the reliance on pre-existing masks in order to generate new images, Costa et al. [17] employed an adversarial autoencoder model to synthesize vessel masks, followed by a Pix2Pix network for mask-to-image translation. Similarly, Gibas et al. [18] used a DCGAN to transform noise space into the vessel segmentation map space and then employed a Pix2Pix (cGAN) for mask-to-image translation. These approaches involve a complex two-step process for controlling the vessel structure of the synthesized images, requiring the use of an additional deep generative model based on adversarial learning alongside GAN-based image-to-image translator training. Additionally, the quality of the synthesized masks directly impacts the quality of the synthesized images. Consequently, if the masks are of insufficient quality, controllable synthesis results in low-quality images. Moreover, pairs of images and their corresponding masks obtained by the U-Net are always utilized to train the Pix2Pix network.

### 2.3. GAN-based DR image synthesis

Numerous researchers have focused on synthesizing retinal images related to DR using GANs. Balasubramanian et al. [24] utilized a DCGAN network to synthesize DR images and performed DR grading using an augmented dataset containing the synthesized images. However, their synthesis was limited to the proliferative class, neglecting other classes. Lim et al. [25] synthesized images specifically for referable DR and vision-threatening DR classes using a style-based generator and a ResNet-based discriminator. Although they analyzed synthesized images for diagnosing these classes, they did not explore the impact of synthesized images on overall DR grading. Furthermore, these studies lacked control over the appearance of DR-related lesions in the generated images, limiting diversity. Zhao et al. [26] introduced the Tub-GAN model, employing style transfer to enhance image diversity, but the lesion information was unclear. Niu et al. [27] utilized vessel segmentation masks and pathological descriptors extracted from reference images to synthesize DR images with controllable lesion characteristics. However, while they manipulated lesion position and quantity, they did not assess the gradability of the synthesized images for the grading task. In contrast, Zhou et al. [28] introduced the DR-GAN model, which synthesized images controllable by the conditional structure of the network, including grading and lesion information. Their synthesized images were evaluated in the grading task. However, the quality of generated image significantly decreased when the input image lacked a lesion mask. Moreover, lesion manipulation relied on the availability of lesion masks, necessitating pre-training of the segmentation model for mask extraction.

### 2.4. Latent space-based image manipulation

Research has shown that GAN-like architectures and their variations can encode various semantics within the latent space for image manipulation[29], [30], [31]. To identify semantics, supervised learning methods often utilize a technique involving the random sampling of a large number of latent codes. The corresponding images are subsequently synthesized and annotated based on predefined image features using a pre-trained classifier [31], [32]**.** However, this manual labeling process is time-consuming and error-prone. At times, determining labels for certain features may not be straightforward, resulting in undiscovered directions due to missing labels. Furthermore, the manual labeling process is conducted by humans, introducing bias into the entire procedure.

In contrast, unsupervised methods offer more flexibility and efficiency. Collins et al. [33] introduced an approach for local image editing that utilizes spherical k-means clustering on activations of the synthesis network of StyleGAN and discovered a disentanglement of semantic objects. Despite its effectiveness, this method requires time-consuming optimization. Another technique, known as GANSpace [34], identifies significant latent directions based on the Principal Component Analysis (PCA) applied in the latent or feature space. These principal components correspond to specific attributes, enabling control over various image features. However, GANSpace relies on extensive data sampling to compute PCA, involving multiple random latent vector samples. A recent advancement in unsupervised latent space editing, SeFa, eliminates the need for training on a generator or extensive data sampling. SeFa leverages the eigenvectors of any GAN convolutional generator block weights, significantly reducing editing time for latent code-based image manipulation.

The current study simplifies the control of vessel structures and lesion areas in synthesized images by leveraging the rich semantic knowledge encoded in the latent space of the network to manipulate synthetic images generated conditioned on different DR grades. Without relying on any feature masks or pre-training of auxiliary networks, we apply the SeFa algorithm to the pre-trained generator of the conditional StyleGAN, to controllably synthesize a broader range of diverse images. In addition, we assess the generated images in both DR detection and grading tasks.

## 3. Methods

Our objective is to solve the problem of learning to controllably synthesize DR fundus images by investigating only a pre-trained GAN and its latent space, aiming to achieve comprehensive control over key features including DR severity grade, lesion type, and vascular structures. This leads us to the ultimate goal of performance improvement in DR diagnosis and grading models by utilizing the diverse synthesized DR images. To achieve this goal, our solution first involves the conditional synthesis of DR images with the desired grade. Second, to introduce greater diversity in the generated images, we unsupervisedly identify semantically meaningful concepts encoded in the latent space. These concepts are then utilized to manipulate specific image features. Finally, the synthesized images from both steps are combined with real data for training and performance improvement of CNN-based classifier for DR analysis. Subsequent sections detail the framework, including dataset description, data preprocessing steps, and the proposed approach.

### 3.1. Data description

In this study, we used the APTOS-2019 dataset [35] from the Kaggle competition for diabetic retinopathy detection provided by the Asia Pacific Tele-Ophthalmology Society. This dataset consists of 3,662 fundus images collected under different imaging conditions acquired at Aravind Eye Hospital in India. The dataset annotation was conducted based on the International Clinical Diabetic Retinopathy scale. Each image was labeled with a severity grade of diabetic retinopathy ranging from 0 to 4 indicating respectively the five categories of normal, mild DR, moderate DR, severe DR, and proliferative DR. Fig.1 illustrates the dataset distribution, as well as the DR images belonging to each of the five classes. From the figure, class distribution of the APTOS dataset is observed to be highly imbalanced in which the percentage of the images belonging to the No DR, Mild DR, Moderate DR, Severe DR, and Proliferative DR classes are almost 49.3%, 10.1%, 27.3%, 5.3%, and 8.06% respectively.

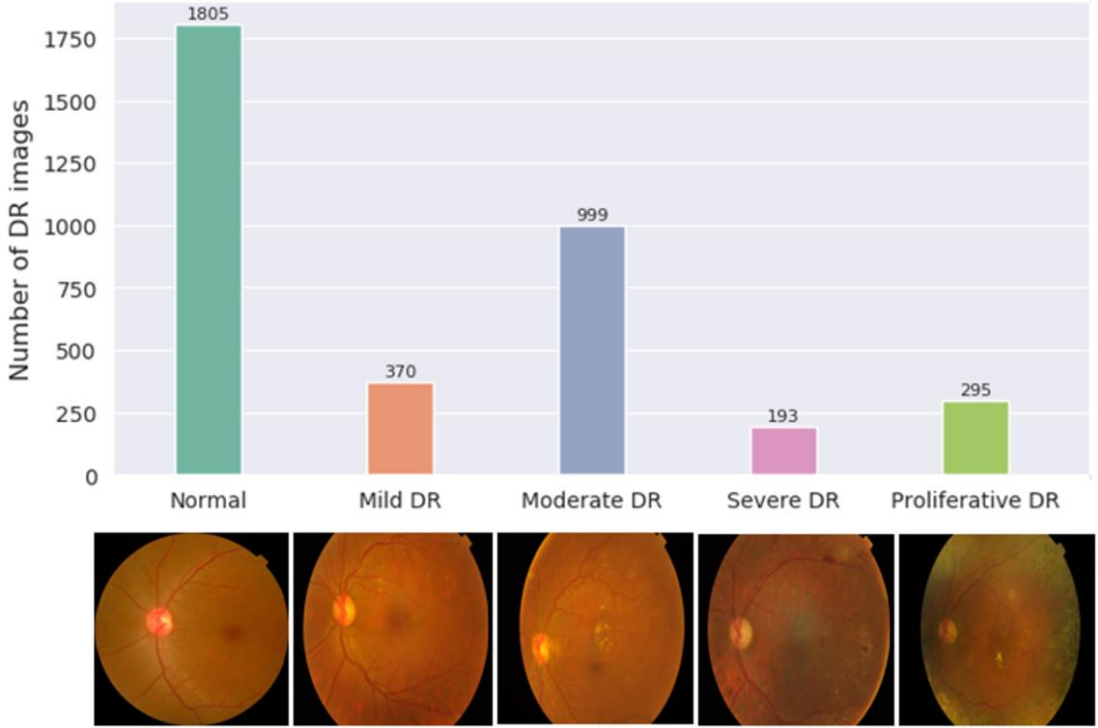

Figure 1: APTOS-2019 dataset used for retinal fundus image synthesis.

### 3.2. Data pre-processing

The dataset comprises images collected from various clinics using different devices, resulting in significant intensity variation. To optimize the training process, we conducted a series of pre-processing steps. Initially, we resized images to a standard size of 256×256 using bilinear interpolation. Subsequently, we implemented "Min-pooling pre-processing", a technique introduced by Graham, to enhance the clarity of blood vessels and lesion areas in fundus images. This involved eliminating black pixels from the background and applying min-pooling filtering to normalize the images with specific parameters as given by:

$$I_c = \alpha I + \beta G(\rho) * I + \gamma \qquad (1)$$

where $I$ denotes input image, $G(\rho)$ denotes the Gaussian filter with a standard deviation of $\rho$, and $*$ denotes the convolution operation. And, $\alpha$, $\beta$, and $\gamma$ are pre-defined parameters. Lastly, we normalized the intensity values of all images to [-1, 1] to eliminate feature bias and ensure uniform distribution across the dataset. Fig.2 illustrates the pre-processing results for samples from different classes, demonstrating the efficacy of the applied techniques

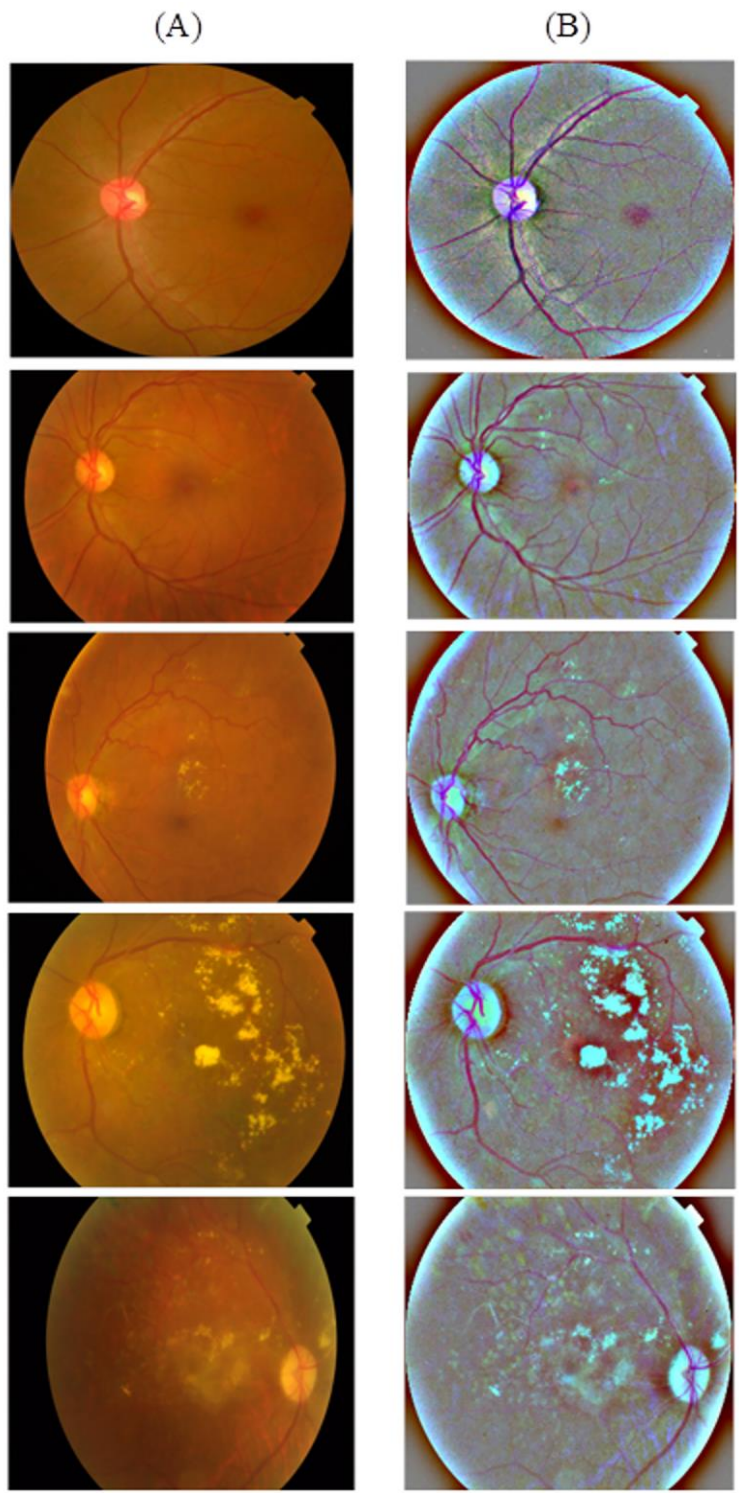

Figure 2: The APTOS dataset preprocessing results. The left and right columns are original retinal images and the preprocessed ones, respectively.

### 3.3. DR fundus synthetic image generation

The first step is to learn DR image synthesis with control over DR grades. To accomplish this, we modify the vanilla StyleGAN model proposed by Karras et al [19]. into a conditional structured GAN to generate retinal fundus images of a desired DR grade. The vanilla StyleGAN features a style-based generator architecture comprising a mapping network and a synthesis network. Instead of directly feeding a latent vector to the generator through an input layer, the feed-forward structured mapping network $f: Z \to W$ takes a latent vector $\mathbf{z}$ and transforms it into an intermediate latent vector $\boldsymbol{w} \in W$. This provides the possibility of controlling image features through the disentanglement of the latent space $Z$. The convolutional structured synthesis network $g$, which is the main core of the image generation process in the model, starts the process from a learned constant tensor. Simultaneously, the synthesis network performs image generation while being controlled by the intermediate latent vectors and the noise sources. The influence from the intermediate latent vectors enhances the perception of the generated images, as the coarse and fine image features becomes clearly distinguishable for the synthesis network. Since the disentangled intermediate latent space has control over the image styles, effective image style manipulation can be performed. This is called style mixing, in which one can easily mix and match styles from different images to generate new images. On the other hand, the noise sources affect the stochastic variations, which refer to the random aspects of the generated images.

The conditional StyleGAN architecture is shown in Fig.3. In order to incorporate the class conditional information into the generator, firstly, $c \in \{0, 1, 2, 3, 4\}$, which refers to the DR grade, is encoded into the one-hot vector $\boldsymbol{y}$ of length 5, and then by multiplying with the $5 \times 512$ matrix of $\mathbf{R} \sim \mathcal{N}(\mu, \sigma^2)$, it is linearly transformed into a 512-length vector. This vector is then concatenated with the latent vector $\mathbf{z}$ to create the final 1024-length vector, which is then fed into the mapping network $f$. This network consists of 8 fully connected layers with leaky ReLU activation functions. The output 512-length intermediate latent vector $\boldsymbol{w}$ is passed into learned affine transformation modules to produce the style vectors $\boldsymbol{s} = (\boldsymbol{s}_s, \boldsymbol{s}_b)$. These vectors are then fed into each level of the synthesis network $g$. The synthesis network consists of convolutional blocks that contain an upsampling layer and two convolutional layers with $3 \times 3$ kernels each accompanied by adaptive instance normalization (AdaIN) operation, which is defined as:

$$\text{AdaIN}(\mathbf{x}_i, \mathbf{s}) = \mathbf{s}_{s,i} \frac{\mathbf{x}_i - \mu(\mathbf{x}_i)}{\sigma(\mathbf{x}_i)} + \mathbf{s}_{b,i}, \tag{2}$$

where each feature map $\mathbf{x}_i$ is normalized separately, and $\mathbf{s}_{s,i}$ and $\mathbf{s}_{b,i}$ are the scaling and the bias components of the style vector $\boldsymbol{s}$ at level $i$.

To generate stochastic details, each block of the synthesis network is provided with explicit noise inputs, which are single-channel images containing Gaussian noise. These noise images are broadcasted to all feature maps using learned scaling factors and then added after each convolutional layer.

Similarly, the discriminator consists of multiple blocks, each comprising two convolutional layers with $3 \times 3$ kernels and a downsampling layer. The last block begins with a mini-batch standard deviation layer, followed by a convolution layer and two linear layers. The activation function for the first linear layer is the leaky ReLU, while the second one is a completely linear layer. In order to provide the discriminator with the class-conditional information $\boldsymbol{y}$, we followed the approach used in [36]. The last linear layer in the progressive discriminator is a 5-length vector. The vector index associated with the input label is considered as the logits. For this purpose, the inner product between the one-hot vector $\boldsymbol{y}$ and the output of the last linear layer is calculated as the final discriminator output.

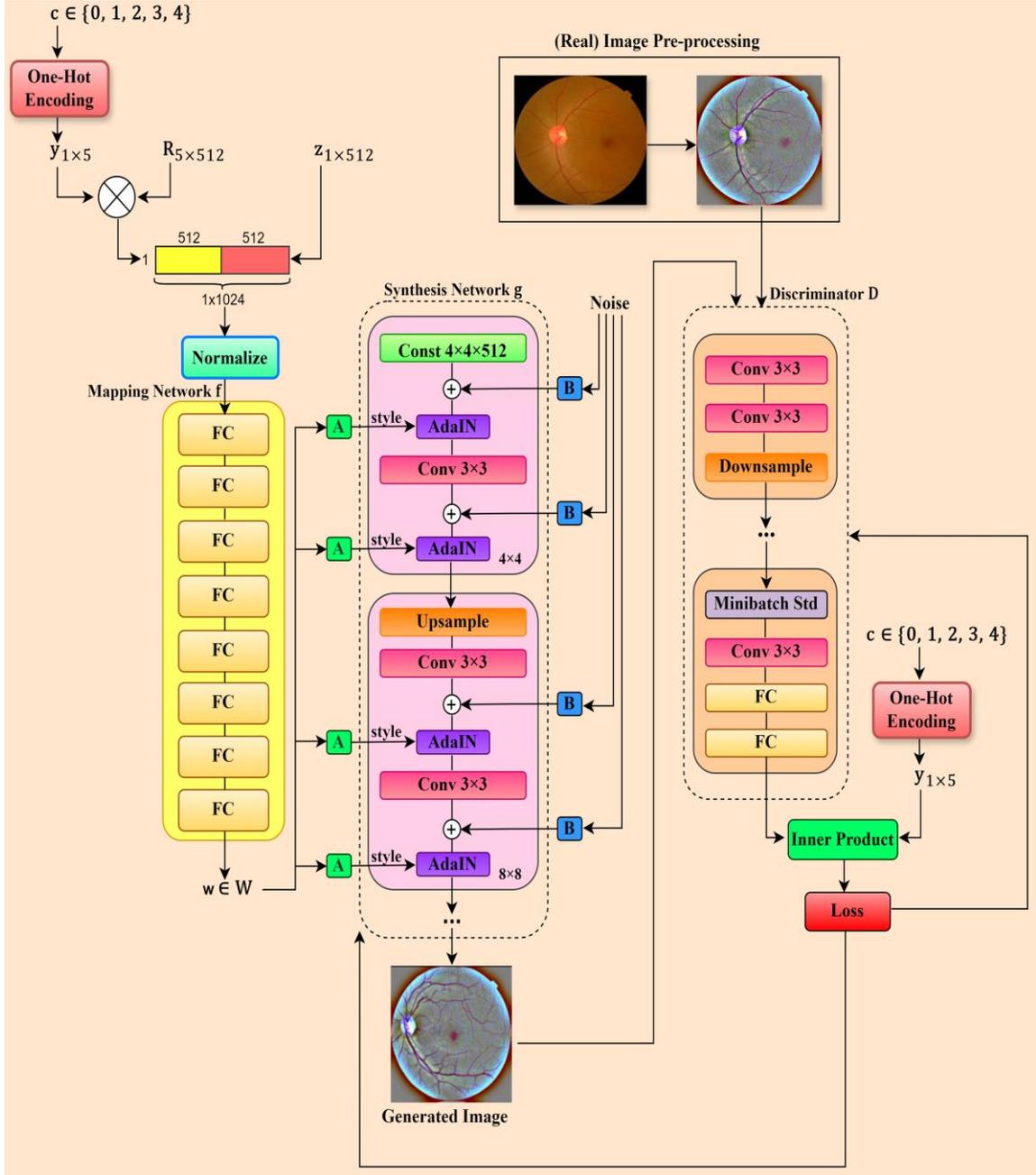

Figure 3: Architecture of the Conditional StyleGAN.

Inspired by [19], we optimize the model using the non-saturating loss [9] with R1 regularization proposed in [36] as defined in Eq. (3):

$$\mathcal{L}_D = -\mathbb{E}_{z \sim p(z)}[log(1 - D(G(z, y)))] - \mathbb{E}_{x \sim p(x)}[log(D(x, y))] + \frac{\gamma}{2}\mathbb{E}_{x \sim p(x)}\left[\|\nabla D_\psi(x, y)\|^2\right],$$
$$\mathcal{L}_G = -\mathbb{E}_{z \sim p(z)}[log(D(G(z, y)))] \tag{3}$$

where G and D are the generator and discriminator respectively, $\boldsymbol{z}$ is a random latent sampled from a prior Gaussian distribution $p(\boldsymbol{z})$, $\boldsymbol{x}$ is a random image sampled from the true data distribution $p(\boldsymbol{x})$, $\boldsymbol{y}$ is the one-hot encoded class label, $\psi$ is the weight of the discriminator and $\gamma$ is set to 10 during training.

### 3.4. Unsupervised identification of DR-related semantic concepts

In the second step, we aim to have more control over the other high-level features in the synthesized images obtained from the first step. After training the GAN model in the previous step, we now explore the latent space of the network, aiming to identify high-level semantic concepts observed in DR images, such as the optic disc, lesions, and vascular structures. By identifying such concepts, we are able to manipulate the high-level features of the synthesized images and generate more diverse images. For this purpose, the recently proposed method SeFa is adopted to identify a set of directions $\boldsymbol{n} \in \mathbb{R}^d$ representing semantically meaningful concepts from the latent space. The interpretable directions $\boldsymbol{n}$ are found in closed-form, through performing eigen-decomposition on the generator weight matrix. Utilizing the semantic knowledge encoded in the latent space $Z \subseteq \mathbb{R}^d$ of GANs, image manipulation can generally be performed through the following equation [34], [37]:

$$edit\ (G(\boldsymbol{z})) = G(\boldsymbol{z}') = G(\boldsymbol{z} + \alpha\boldsymbol{n}) \tag{4}$$

where $\boldsymbol{z} \in Z$ is the input latent code and $\boldsymbol{n} \in \mathbb{R}^d$ denotes a certain direction in the latent space representing a semantic concept. The $edit(.)$ indicates the image manipulation operation, and the manipulation intensity can be adjusted by the parameter $\alpha$. According to the Eq. (4), target semantic modification is performed by linearly moving the latent code $\boldsymbol{z}$ along the identified direction $\boldsymbol{n}$. Regarding the convolutional-based architecture of the generator, each convolutional layer is a transformation from one space to another. The first layer transformation, which directly operates on the latent space, can be considered as an affine transformation denoted as:

$$G_1(\boldsymbol{z}) \triangleq \boldsymbol{y} = \boldsymbol{A}\boldsymbol{z} + \boldsymbol{b}, \tag{5}$$

where the m-dimensional vector $\boldsymbol{y}$ is the projected code. $\boldsymbol{A} \in \mathbb{R}^{m \times d}$ and $\boldsymbol{b} \in \mathbb{R}^m$ represent the weight and bias of $G_1$, respectively. Following the general manipulation model (Eq. (4)), an instance-independent manipulation process can be accomplished based on the first layer affine transformation.

$$\begin{aligned} \boldsymbol{y}' \triangleq G_1(\boldsymbol{z}') &= G_1(\boldsymbol{z} + \alpha\boldsymbol{n}) \\ &= \boldsymbol{A}\boldsymbol{z} + \boldsymbol{b} + \alpha\boldsymbol{A}\boldsymbol{n} = \boldsymbol{y} + \alpha\boldsymbol{A}\boldsymbol{n}. \end{aligned} \tag{6}$$

This equation suggests that the generator weights should contain the essential knowledge of the image variation, as the editing operation is mainly affected by the weight parameter $\boldsymbol{A}$. Therefore, the weight matrix $\boldsymbol{A}$ is decomposed in order to discover the directions $\boldsymbol{n}$ that can cause a significant change in the output image. In order to explore the latent semantics to find the top-$k$ optimal directions $\boldsymbol{N} = [\boldsymbol{n}_1, \boldsymbol{n}_2, \dots, \boldsymbol{n}_k]$, we need to solve the following optimization problem:

$$\boldsymbol{N}^* = \underset{\{\boldsymbol{N} \in \mathbb{R}^{d \times k} : \boldsymbol{n}_i^T \boldsymbol{n}_i = 1\ \forall i=1,\cdots,k\}}{arg\ max} \sum_{i=1}^{k} \|\boldsymbol{A}\boldsymbol{n}_i\|_2^2, \tag{7}$$

To obtain the top-$k$ meaningful directions, columns of $\boldsymbol{N}$ as the eigenvectors of $\boldsymbol{A}^T\boldsymbol{A}$ is selected, which correspond to the $k$ largest eigenvalues.

### 3.5. DR fundus image classification

The DR images generated in the previous two steps are combined with real images to train ResNet50, a well-known CNN-based classifier, assessing its potential benefit from synthesized images to effectively improve DR detection and grading. ResNet was proposed to address the issue of accuracy degradation that occurs as the network becomes deeper. The solution involved introducing identity shortcut connections between the input of the first layer and the output of the last layer in a same residual block. This residual framework facilitates easier optimization of the network, leading to improved accuracy as network depth increases. ResNet50 comprises 50 layers, including a convolutional layer, 16 residual blocks with a total of 48 convolutional layers, and a fully connected layer. In this step, we perform DR diagnosis and grading by utilizing images generated class-conditionally from the first step. Furthermore, images generated by SeFa-based manipulation are used to train the classifier for DR diagnosis. To this end, we implement transfer learning by leveraging the weights of a ResNet50 model pre-trained on the ImageNet dataset. This strategy enables the integration of existing knowledge, thereby enhancing the generalizability of the model. The pre-trained ResNet50 model serves as the feature extractor backbone, which is subsequently coupled with a classification head, as depicted in Fig.4. The classification head encompasses Batch Normalization and Dropout layers, strategically incorporated to mitigate overfitting tendencies. These layers are followed by linear and ReLU layers, forming a cohesive architecture to process and classify the extracted features effectively. Furthermore, we introduce an AdaptiveConcatPool layer, comprising AdaptiveAveragePool and AdaptiveMaxPool layers. Specifically, the outputs of both pooling layers are concatenated, enabling the utilization of information from both layers and thereby preserving the feature representations of the backbone.

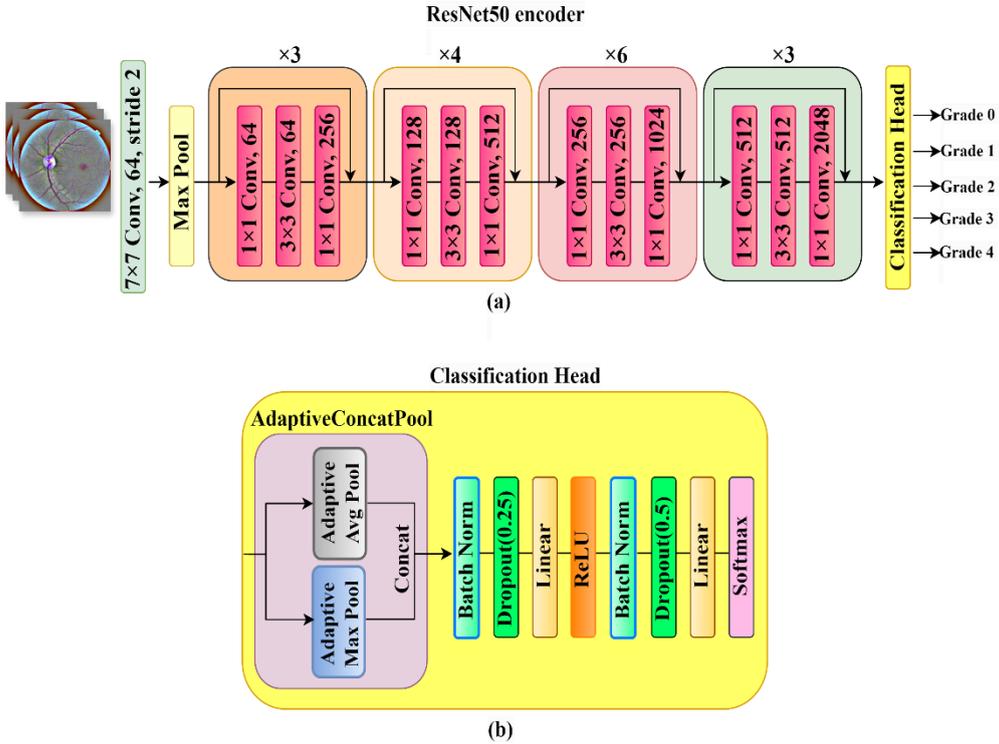

Figure 4: Architecture of the classifier.

## 4. Experiments and results

In this study, we conduct various experiments to quantitatively and qualitatively evaluate the quality of synthetic DR fundus images, whether they are generated directly by the conditional StyleGAN or obtained through the manipulation of previously synthesized images.

### 4.1. Training process

The conditional StyleGAN underwent progressive growth training from a resolution of 4×4 to 256×256, with batch sizes of 128, 128, 128, 64, 32, 16, and 8, respectively. Both the generator and discriminator were optimized using the Adam optimizer, with a learning rate of 1e-8. The training duration for the conditional StyleGAN totaled approximately 59 hours, utilizing a NVIDIA GTX 1080 Ti GPU. Seventy percent of the Aptos 2019 dataset was utilized for training the classifier, while the remaining 30% was split with 20% allocated for validation and 10% for testing purposes. During training, the ResNet50 model underwent fine-tuning, wherein the classification head was trained for a maximum 50 epochs with all other layers frozen. Subsequently, all layers were unfrozen and trained for a maximum of 100 epochs. Training parameters included the use of the Adam optimizer with a learning rate of 3e-3 and a batch size of 64. To mitigate overfitting, the LabelSmoothingCrossEntropy loss function was implemented as a regularization technique. The classifier training was conducted on a NVIDIA Tesla T4 GPU provided by Google Colab, utilizing the Fastai framework [38]. Meanwhile, the conditional StyleGAN and the SeFa method were implemented using Tensorflow and Pytorch, respectively, based on their official implementations.[1]

### 4.2. Qualitative assessments

Qualitative assessments of synthetic images are conducted first for the DR images directly generated by the conditional StyleGAN and then for the semantically manipulated versions resulting from comparative experiments, which will be discussed further.

#### 4.2.1. Conditional diabetic retinopathy generation

Fig.5 presents three samples per class generated by the conditional StyleGAN. As can be observed, the generated images exhibit a remarkable level of quality and realism. Their realistic appearance is evident in the accurate portrayal of anatomical structures, including the optic disc, vessels, and lesions specialized for each class, showcasing a high degree of fidelity. Furthermore, the diversity and variability observed in the generated images underscore the versatility of the conditional StyleGAN in synthesizing various manifestations of DR.

#### 4.2.2. Semantic manipulation of synthetic diabetic retinopathy images

In this section, firstly, synthesized images are edited using semantically meaningful directions computed in the latent space by the SeFa algorithm. In other words, image manipulation is performed by controlling the lesions, vessel structure, and other attributes. Next, style-based editing experiment is conducted to visually analyze and compare the controllability of the attributes and the semantic manipulability of the images using the pre-trained style-based generator, with and without utilizing the SeFa approach. In these experiments, a subset of layers, categorized as bottom layers, middle layers, and top layers, are experimented with to be interpreted. Bottom layers refer to layers 0-1, corresponding to coarse spatial resolution of 4×4. Middle layers refer to layers 2-5, corresponding to middle resolutions of 8×8 and 16×16. Top layers refer to layers 6-13, corresponding to fine resolutions of 32×32, 64×64, 128×128, and 256×256.

---

[1] https://github.com/NVlabs/stylegan
http://github.com/genforce/sefa

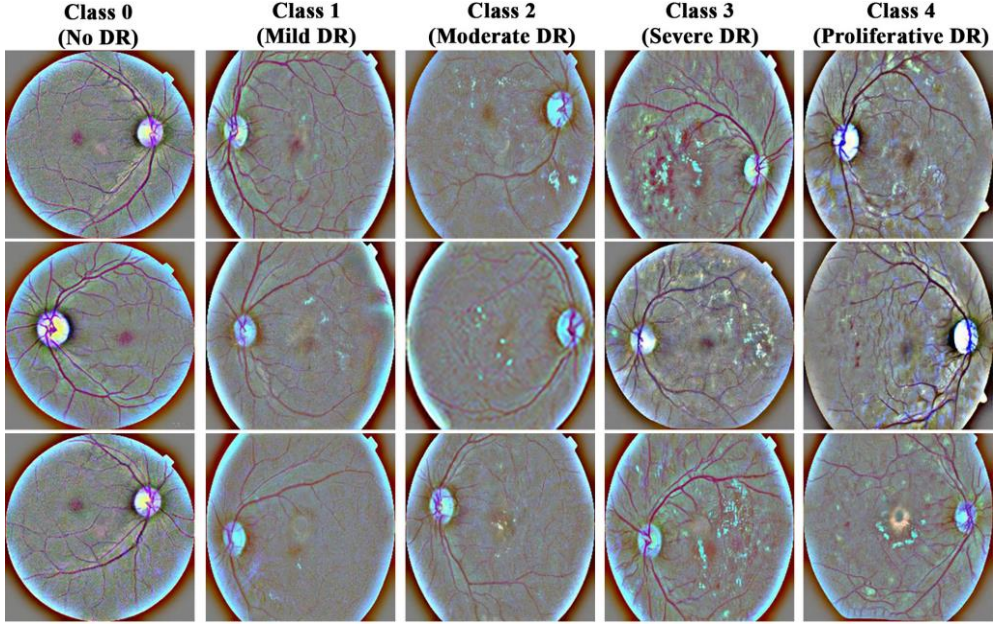

Figure 5: Three samples per class generated by the conditional StyleGAN.

- **SeFa-based manipulation**

As can be seen in Fig.6, the bottom layers are responsible for generating coarse attributes, such as determining the location of the optic disc to the right or left and whether retinal images are complete or incomplete in a circular shape. In fact, modifying the latent code using the knowledge from these layers reveals changes in the horizontal location of the optic disc and the overall appearance of the retinal images. Although these changes are not limited solely to this semantic, and as can be seen, other attributes such as the vertical location of the optic disc, general shape of vessels, lesions, etc., also undergo minor changes. However, by traversing the latent code along the discovered direction, a semantically obvious continuous transition in the mentioned attributes can be observed. Notably, this semantic is not controlled by the middle and top layers. In other words, varying the latent code using the knowledge of the middle and top layers does not impact the horizontal location of the optic disc and the overall appearance of the retinal images, but it does affect other attributes, which will be discussed further. The middle layers tend to control the concentration of vessels. Specifically, moving along the identified direction based on these layers, as can be seen in the right images, decreases the concentration of vessels, and fewer vessels branching into finer vessels can be observed compared to the original. Conversely, moving away from the direction, as illustrated in the left images, leads to a higher vessel concentration and a tendency to generate more branches. Similar to the observation in the bottom layers, manipulating the latent code using the knowledge from the middle layers does not solely change the corresponding semantic, i.e., the concentration of vessels, and there is also a negligible change in the lesion structures. However, the horizontal location of the optic disc remains unaltered. The top layers determine disease-related lesions in terms of their location and intensity. When moving along the interpretable direction, as observed in the right images, the output images exhibit increased lesion intensity. Conversely, moving away from this direction tends to remove lesions, resulting in specific control over lesion severity, as shown in the left images. Once again, our observation confirms that the lesion attribute is not the only one altered by the top layers. However, what allows one to determine that these layers primarily correspond to this semantic is that, by traversing through the direction, an obvious continuous transition in lesion intensity can be observed. This transition is not observed in the middle layers. Fig.7 shows detailed steps of the mentioned coarse attributes manipulation and Fig.8 and 9 visualize detailed steps of vessel and lesion manipulation, respectively, for the same samples.

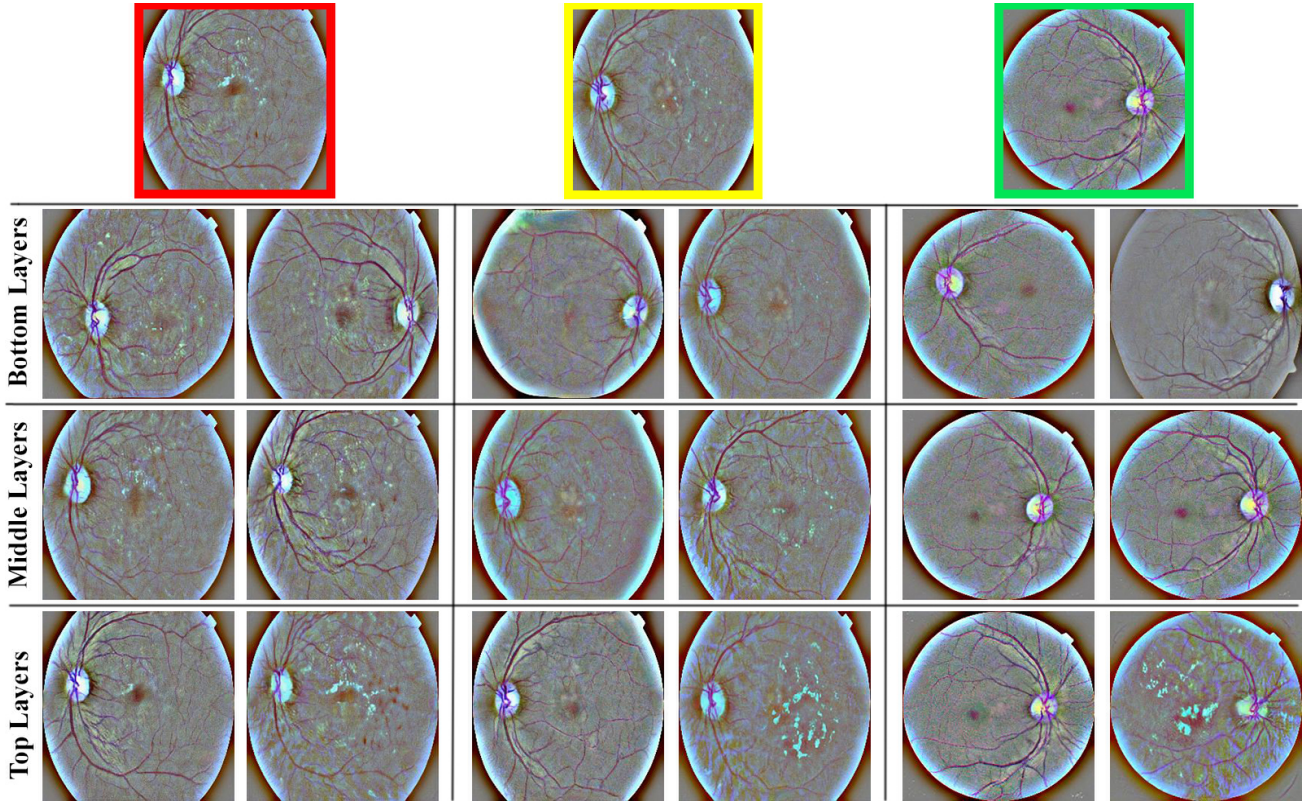

Figure 6: Image editing using discovered interpretable directions. The topmost images are the original samples generated by the conditional StyleGAN. The images with red, yellow, and green outlines are from class 3, class1, and class 0, respectively. Images in the lower rows show the manipulation results based on different layers. In each row, the right image depicts the manipulation when moving along the direction, while the left image shows the result when moving away from the direction.

- **Style-based manipulation**

  The style-based generator architecture enables the manipulation of generated images by modifying styles at different levels. Specifically, first, the mapping network outputs $\mathbf{w}_1$, $\mathbf{w}_2$ corresponding to two latent codes $\mathbf{z}_1$, $\mathbf{z}_2$, which are then transformed into corresponding style vectors through affine transformations. Subsequently, the synthesis network generates a new image by blending the produced styles, i.e., applying one style to a subset of layers while the other style controls the rest of the layers. In fact, as each style influences distinct blocks of the network, altering particular sets of styles selectively affects specific attributes of the generated image.

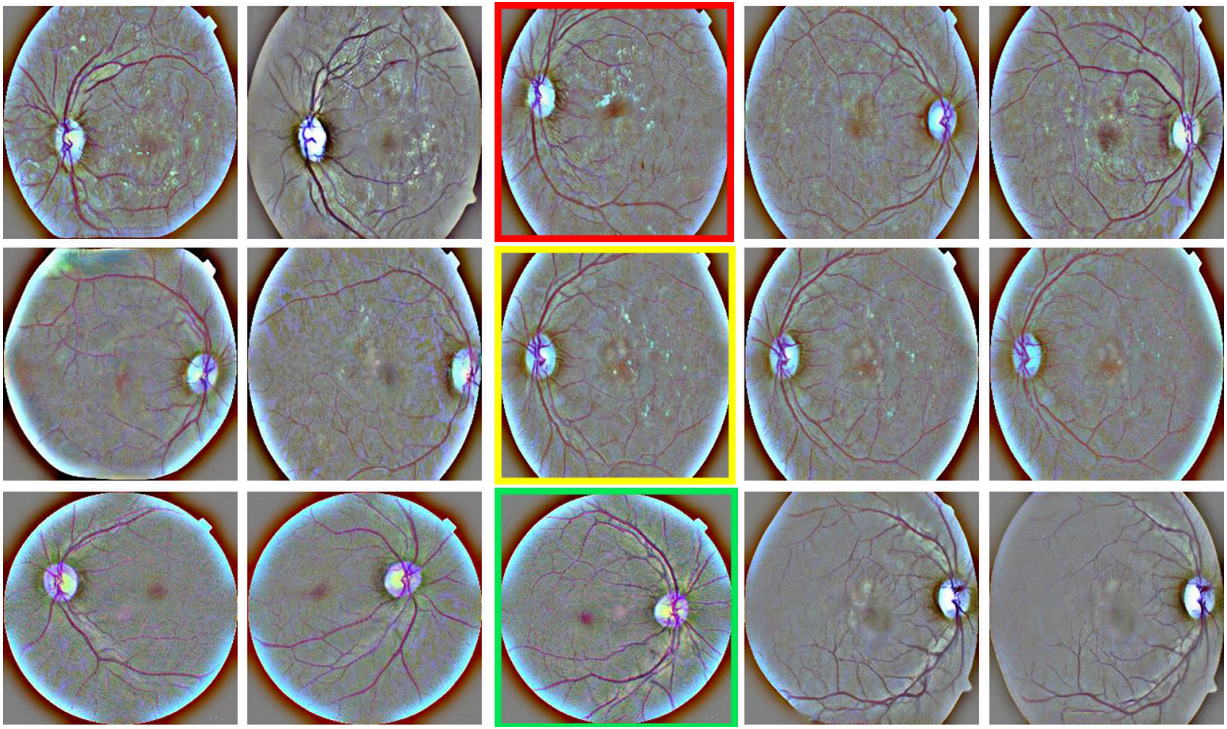

Figure 7: Detailed steps for manipulating course attributes, including the optic disc location and the overall image shape, for the three samples.

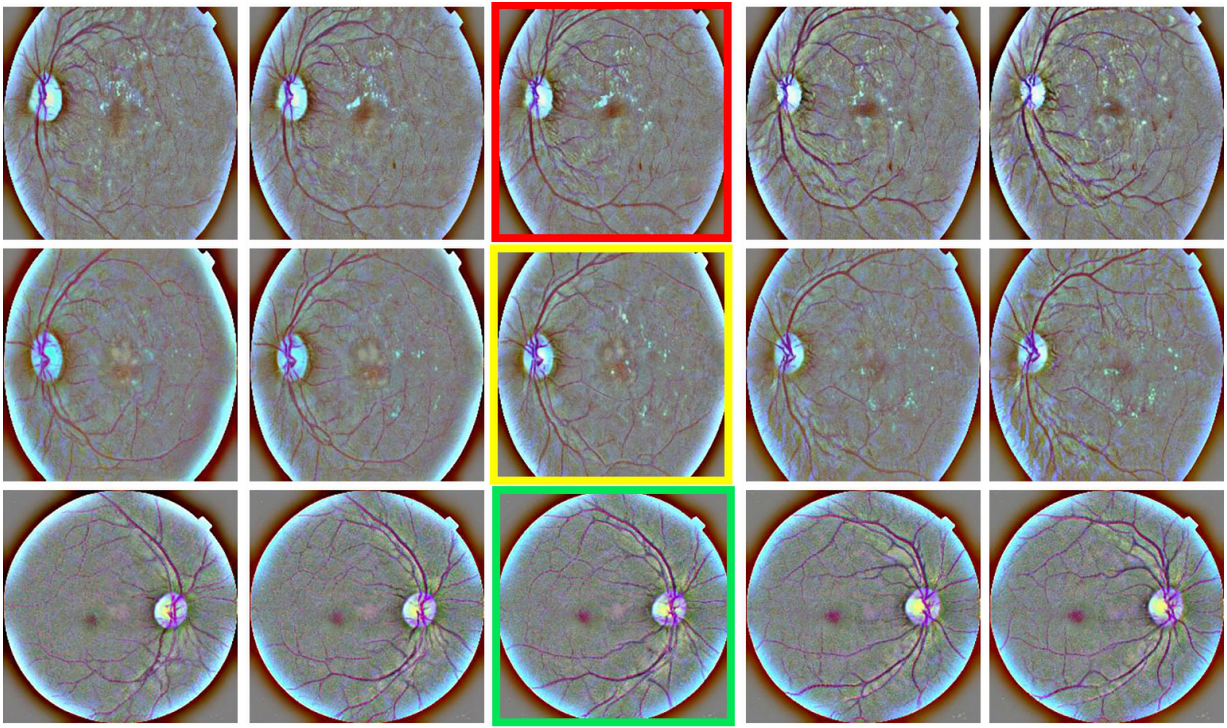

Figure 8: Detailed steps of vessel manipulation for the three samples.

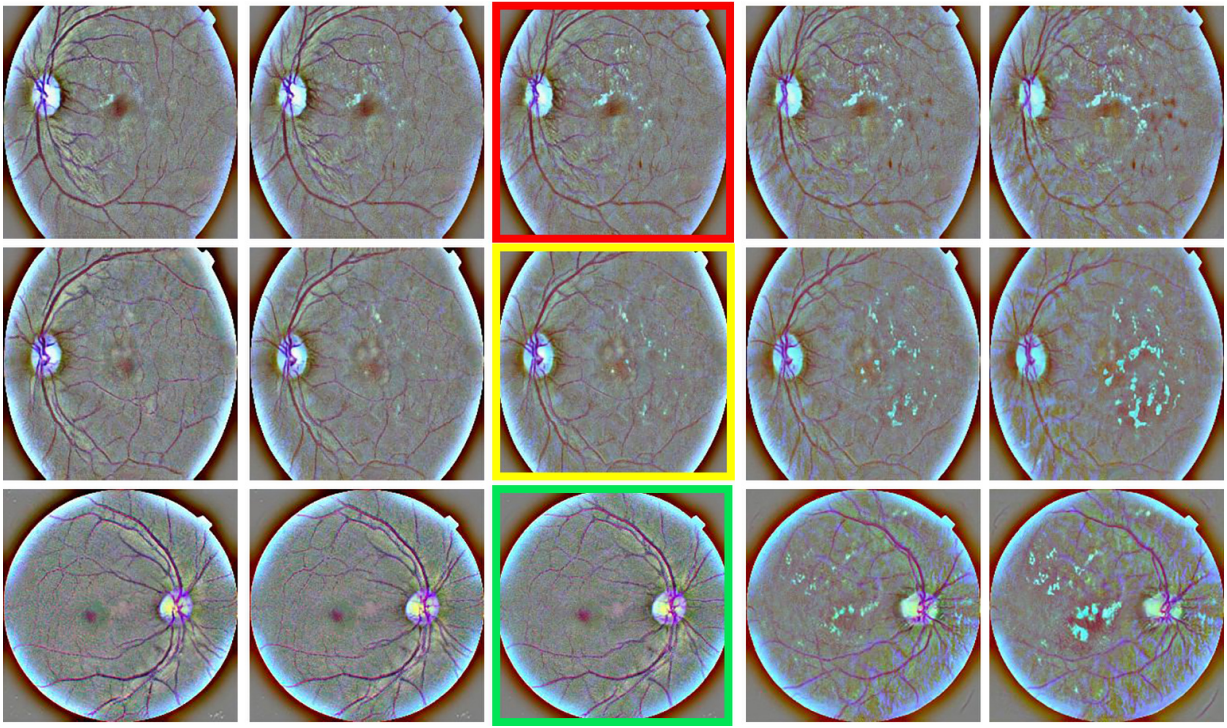

Figure 9: Detailed steps of lesion manipulation for the three samples.

The Fig.10 shows examples of images created by combining two different style codes at different levels. Two groups of images, referred to as source A and source B, as demonstrated in the leftmost column and the top row of the Fig.10, are synthesized from class 0 and class 4, respectively, in order to generate the maximum possible change by combining the attributes of both groups. The images in the first three rows are generated by applying a specific subset of styles from source B to the bottom layers, while the remainder is selected from source A. As can be observed, the styles corresponding to the bottom layers control semantic high-level attributes, such as the overall appearance of the retinal images, whether they are complete or incomplete circular images, the position of the optical disk (right or left), and the general structure of the vessel network. The images in the last row are generated by copying the styles corresponding to the top layers from source B. This does not alter the mentioned coarse attributes of the images from source A; however, it introduces lesions from source B. These two observations align with the findings of SeFa, which demonstrated that the bottom layers determine the horizontal location of the optic disc and the overall appearance of the retinal images, while the top layers influence the appearance of lesions. However, employing SeFa offers us much greater control over the attributes of the synthesized images. Finally, the observations based on the middle layers show no apparent manipulation. However, when employing the SeFa approach for image manipulation based on these layers, it effectively controls the structure of vessel networks.

*4.3. Quantitative assessments*

This section presents a quantitative evaluation of controllably synthesized images through various experiments in grading and detection tasks, showcasing the quality and diversity of images synthesized by conditional StyleGAN. Furthermore, we introduce a novel, effective SeFa-based data augmentation strategy, specifically tailored for the detection task. This augmentation technique further boosts classifier performance by effectively manipulating synthesized images. In addition, we juxtapose the results obtained in both grading and detection tasks, with existing literature, employing varied evaluation metrics, thereby highlighting the superiority of our proposed approach over state-of-the-art methods.

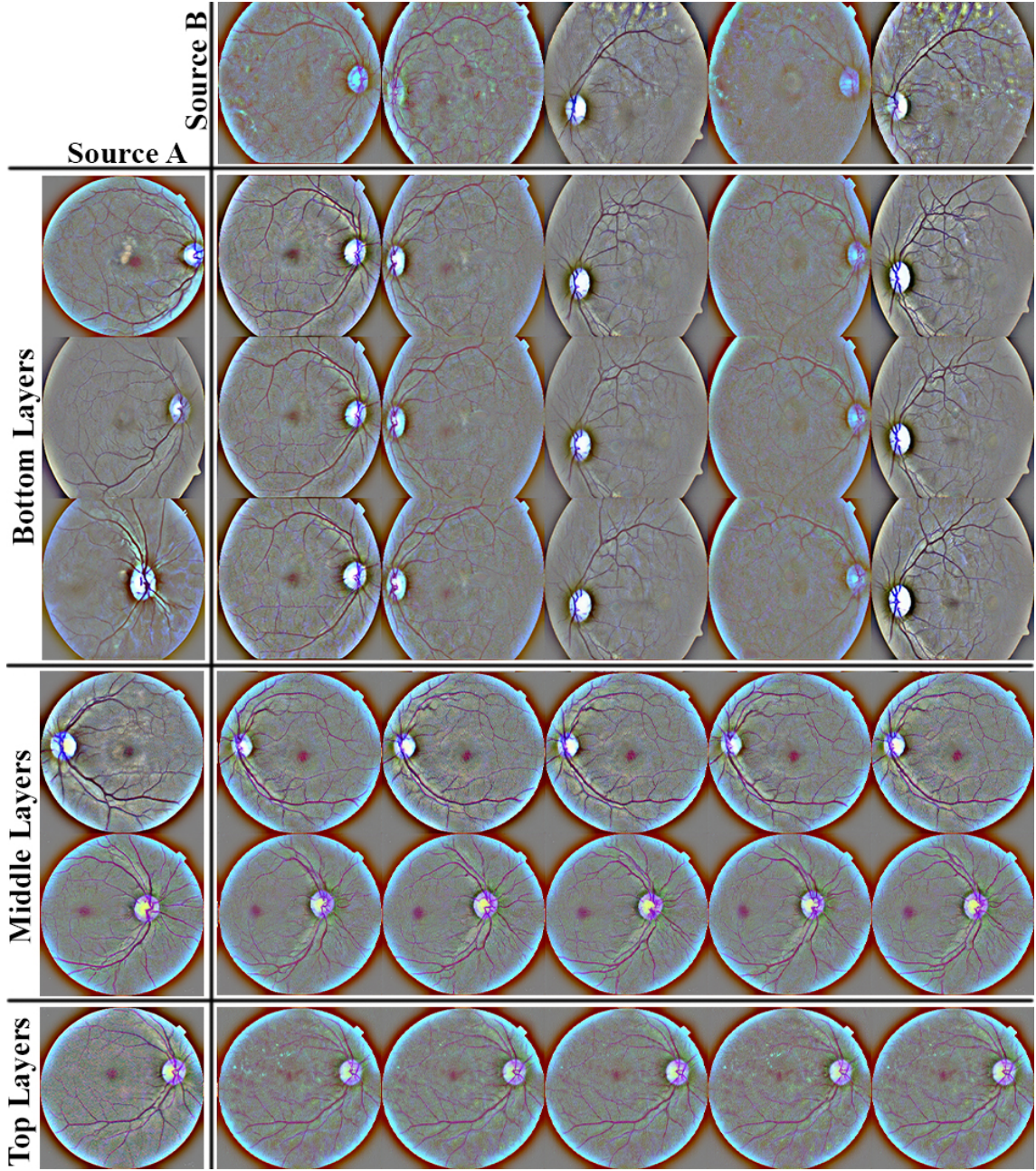

Figure 10: Images of source A are from the normal class, while images of source B are from the proliferative class. The remaining images are generated by substituting a subset of styles from source B for the corresponding styles from source A. The bottom layers, middle layers, and top layers indicate resolutions of $4^2$, $8^2 - 16^2$, and $32^2 - 256^2$, respectively.

Each comparative experiment employs a fixed set of real images in both validation and test datasets, ensuring a fair comparison. With the exception of the effective SeFa-based data augmentation experiment, the class distribution of synthesized data aligns entirely with that of the real training data in all experiments. Consequently, the setting still remains challenging due to the unbalanced mixed dataset. In the case of SeFa-based data augmentation, an equal number of synthesized data from each class is combined with the existing unbalanced real images, ultimately balancing the total number of abnormal and normal images for binary classification. Fig.11 and 12 illustrate the classification with the data augmented using only conditional StyleGAN and the effective SeFa-based data augmentation, respectively.

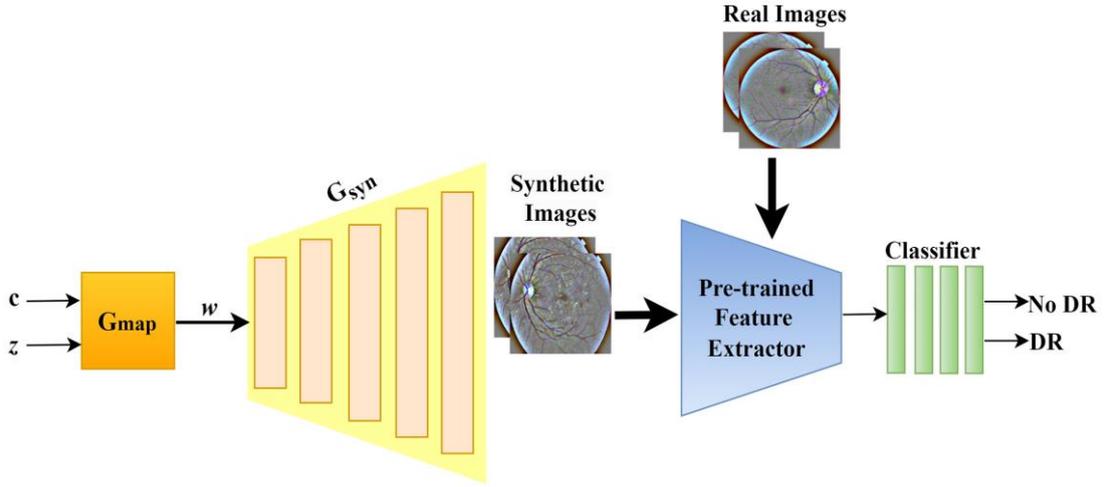

Figure 11: Classification with the data augmented using only conditional StyleGAN.

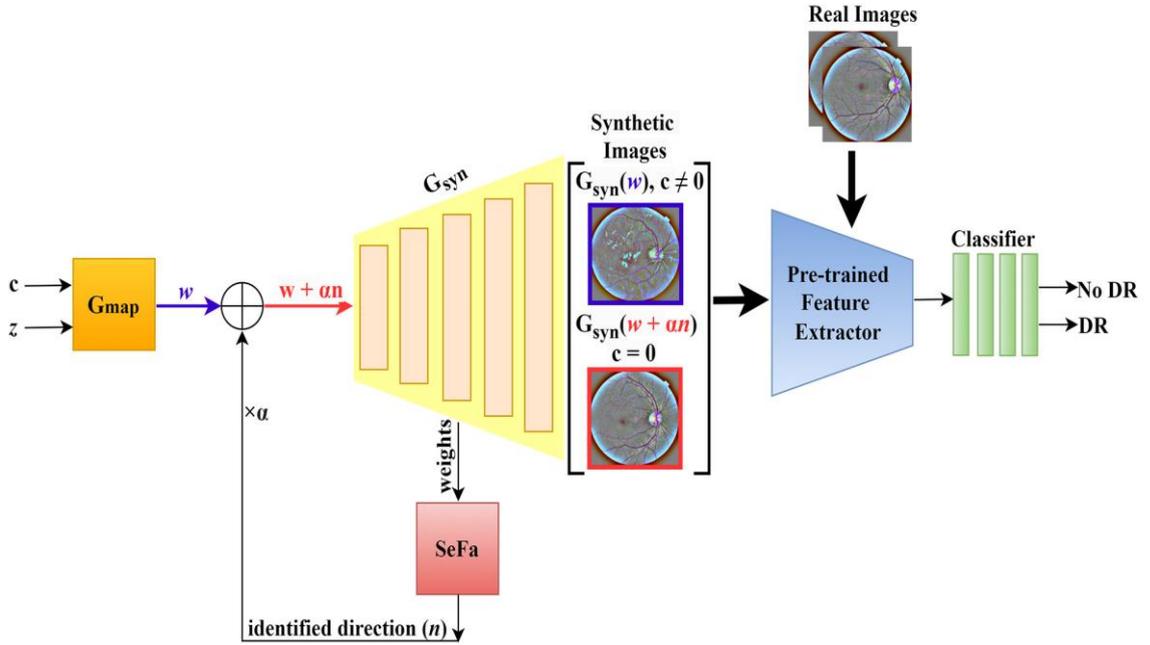

Figure 12: Classification with the effective SeFa-based data augmentation.

*4.3.1. Evaluation metrics*

Classification performance was evaluated using a comprehensive set of criteria, including accuracy, recall or sensitivity, specificity, precision, F1-score, and Quadratic Weighted Kappa (QWK) score. The total number of correctly classified data instances divided by the total number of data instances is known as accuracy. Recall or sensitivity demonstrates the ability of the model to correctly identify all positive instances. It is mathematically calculated as the number of accurately categorized positive instances divided by the total number of actual positive instances. Specificity is defined as the percentage of negative instances that are correctly predicted as negative. Precision is the number of correctly classified positive instances divided by the total number of instances predicted as positive. This shows what proportion of positive predictions were actually correct. F1-score, which is defined as the harmonic mean of precision and recall, addresses the trade-off between precision and recall by providing a balanced measure of both precision and recall. QWK measures the agreement or disagreement between the model's prediction and the ground truth. This score ranges from -1 through 0 to 1, which means total disagreement, random classification, and total agreement, respectively [39]. These metrics are mathematically computed using the following equations:

$$\text{Accuracy} = \frac{TP + TN}{TP + FP + TN + FN} \qquad (8)$$

$$Recall = sensitivity = \frac{TP}{TP + FN} \qquad (9)$$

$$specificity = \frac{TN}{TN + FP} \qquad (10)$$

$$Pecision = \frac{TP}{TP + FP} \qquad (11)$$

$$F1 = 2 * \frac{Precision * Recall}{Precision + Recall} \qquad (12)$$

$$QWK = 1 - \frac{\sum_{i,j} W_{ij} O_{ij}}{\sum_{i,j} W_{ij} E_{ij}} \qquad (13)$$

where TP, FP, TN, and FN refer to the number of true positives, false positives, true negatives, and false negatives, respectively. In the QWK formula, *N* is the number of classes, *i* and *j* are the indices indicating the true class and the predicted class, respectively. The confusion matrix is denoted by O, the expected matrix by E, and the weight matrix by W, defined by the following equation:

$$W_{ij} = 1 - \frac{(i-j)^2}{(N-1)^2} \tag{14}$$

In addition, the Receiver Operating Characteristic (ROC) curve is plotted, illustrating the relationship between the true positive rate and the false positive rate. The Area Under the Curve (AUC), representing the total area beneath this curve, is also calculated.

*4.3.2. DR severity grading*

Here, we present a series of six comparative experiments conducted to quantitatively evaluate the synthesized DR images specifically in the DR severity grading task. The experimental results, as well as the comparison between our best result and some of the state-of-the-art methods, are shown in Tables 1 and 2 respectively, and the conducted comparative experiments are explained bellow.

a. **Real:** The classifier is trained on only the real images without employing any classical data augmentation. This experiment is considered as the baseline classification.

b. **Synthetic:** The classifier is trained on only the generated images without using any real images nor any classical data augmentation. This provides a direct quality evaluation of the synthetic data.

c. **Real + Synthetic (1):** A combination of synthetic and real data is used for training where the size of data in each category is halved. Therefore, the size of the synthetic data becomes half the size of the real data.

d. **Real + Synthetic (2):** A combination of synthetic and real data with the same proportion is used to train the classifier.

e. **Real + Synthetic (3):** A combination of synthetic and real data is used for training where the size of data in each category is doubled. Therefore, the size of the synthetic data becomes twice the size of the real data.

f. **Synthetic → Real:** The classifier is first pre-trained on the synthesized images used in experiment (b) and then is fine-tuned on the real data utilized in experiment (a).

The experimental results demonstrated in the Table 1 show that expanding existing real dataset by adding generated images generally improves the classifier performance significantly. Specifically, when training the classifier on a mix of real data and synthetic data with the same ratio, this improvement is slightly better compared to using a lower or higher ratio, leading to enhancements of 4.91%, 7.86%, 15.98%, 1.46%, 5.07%, 14.76%, and 1.29% in accuracy, QWK, sensitivity, specificity, precision, F1-score, and AUC-ROC, respectively. Moreover, pre-training the classifier on generated images followed by fine-tuning with real ones leads to the greatest improvement over the baseline. Specifically, accuracy, QWK, sensitivity, specificity, precision, F1-score, and AUC-ROC are enhanced by 7%, 10.61%, 18.9%, 1.7%, 16.27%, 20.92%, and 3.68%, respectively, with respect to the case of training the classifier only on the real data. These results indicate that the generated images are adequately diverse and realistic to be beneficial for further performance improvements.

Table 1: Performance comparison of the ResNet50 trained on different training sets for DR severity classification. The metrics are calculated based on macro average.

| Experiments | Accuracy | QWK | Sensitivity | Specificity | Precision | F1-score | AUC-ROC |
|---|---|---|---|---|---|---|---|
| Real | 0.7787 | 0.7923 | 0.5425 | 0.9407 | 0.6213 | 0.5508 | 0.8854 |
| Synthetic | 0.6366 | 0.6411 | 0.4612 | 0.9057 | 0.4588 | 0.4555 | 0.7870 |
| Real + Synthetic(1) | 0.8060 | 0.8484 | 0.5970 | 0.9488 | <u>0.6809</u> | 0.6172 | <u>0.9070</u> |
| Real + Synthetic(2) | <u>0.8169</u> | <u>0.8546</u> | <u>0.6292</u> | <u>0.9544</u> | 0.6528 | <u>0.6321</u> | 0.8968 |
| Real + Synthetic(3) | 0.8087 | 0.8469 | 0.6224 | 0.9511 | 0.6543 | 0.6280 | 0.8893 |
| Synthetic → Real | **0.8333** | **0.8764** | **0.6450** | **0.9567** | **0.7224** | **0.6660** | **0.9180** |

Table 2 shows the best performance of our classifier, achieved through fine-tuning with real data after pre-training on synthetic data, alongside some other related methods conducted for the DR severity classification task. From the table, our classifier surpasses other related models. Specifically, the proposed approach attains an accuracy of 83.33% and a specificity of 95.67%, significantly greater than the corresponding values obtained with Inception V3 [40]. Moreover, in terms of accuracy, our model slightly outperforms the CNN299 [41] model, which extensively performed classical data augmentation to increase the size of dataset by 20 times. In addition, our ResNet50 classifier outperforms the ResNet50 [42] model, which utilized classical data augmentation during training, while our classifier utilizes GAN-based data augmentation. Notably, the proposed approach enhances the accuracy, QWK, F1-score, and AUC-ROC of the ResNet50 [42] model by 4.03%, 25.07%, 2.83%, and 6.74, respectively. This demonstrates the increased diversity that synthetic data augmentation potentially introduces to the existing dataset. Comparing the computed metrics with the DenseNet169 + CBAM [43] also indicates that a ResNet50 utilizing synthetically GAN-based data augmentation achieves superior overall performance compared to a much deeper network modified with an attention module. Furthermore, our proposed approach achieves higher scores in comparison with the pre-trained models VGG16 [44], Xception [44], and MobileNet [44]. When these models are combined together with the other four pre-trained deep classifiers to create the self-adaptive ensemble model [44], the model is still outperformed by ours in terms of accuracy and precision, while the obtained values for recall and QWK are very close. Finally, the ECOC-SVM model [45], which is also designed based on the ensemble of two deep models of ResNet18 and ShuffleNet, is outperformed by our single model across most metrics. Fig.13 illustrates the class-wise performance superiority of the two best models trained on GAN-based augmented data over the baseline classifier through the confusion matrices. As can be observed, expanding the existing real dataset by augmenting synthetic data with the same ratio improves the detection of all the abnormal classes by 20% for the class 1, 2.5% for the class 2, 100% for the class 3, and 66.7% for class 4. In addition, pre-training the classifier on generated images followed by fine-tuning with real ones further improves the class-wise detection accuracies specifically by 10% for class 2 and 150% for class 3 over the baseline.

Table 2: Comparison of the ResNet50 Performance with related methods for severity classification of DR using APTOS 2019 dataset. The metrics are calculated based on macro average.

| Model | Accuracy | QWK | Sensitivity | Specificity | Precision | F1-score | AUC-ROC |
|---|---|---|---|---|---|---|---|
| Hybrid Inception-ResNet-V2 [46] | 0.8218 | - | - | - | - | - | - |
| Inception V3 [40] | 0.787 | - | 0.636 | 0.853 | - | - | - |
| ResNet50 [42] | 0.8010 | 0.7007 | - | - | - | 0.6477 | 0.86 |
| CNN299 [41] | 0.832 | - | - | - | - | - | - |
| DenseNet169 + CBAM [43] | 0.81 | 0.8607 | - | - | - | 0.60 | - |
| ECOC-SVM [45] | 0.806 | - | **0.646** | 0.952 | 0.6267 | 0.633 | - |
| VGG16 [44] | 0.8035 | 0.8651 | 0.5767 | - | 0.6549 | 0.5989 | 0.9146 |
| Xception [44] | 0.7804 | 0.8177 | 0.5056 | - | 0.6589 | 0.5203 | 0.8907 |
| MobileNet [44] | 0.7954 | 0.8557 | 0.6108 | - | 0.6599 | 0.6297 | 0.9073 |
| Ours | **0.8333** | **0.8764** | <u>0.6450</u> | **0.9567** | **0.7224** | **0.6660** | **0.9180** |

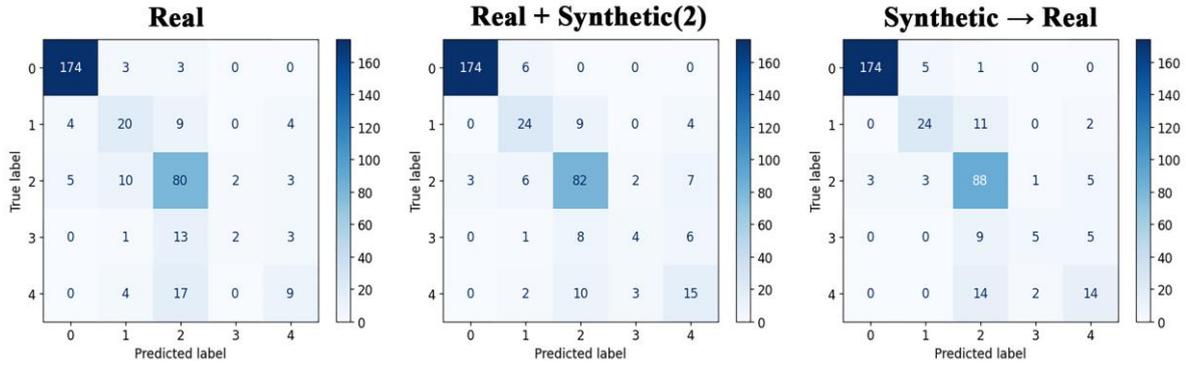

Figure 13: Confusion matrices for DR severity grading conducted across different experiments.

Fig.14 depicts the micro- and macro-averaged ROC curves of the classifier for DR severity grading across two experiments: the baseline (a) and the experiment yielding the best results (f). Analysis of the figure reveals that fine-tuning the model with real images subsequent to pre-training on synthetic ones results in a superior AUC score for ROC, indicative of enhanced discriminatory capacity across different DR grades, compared to training the model solely with real images.

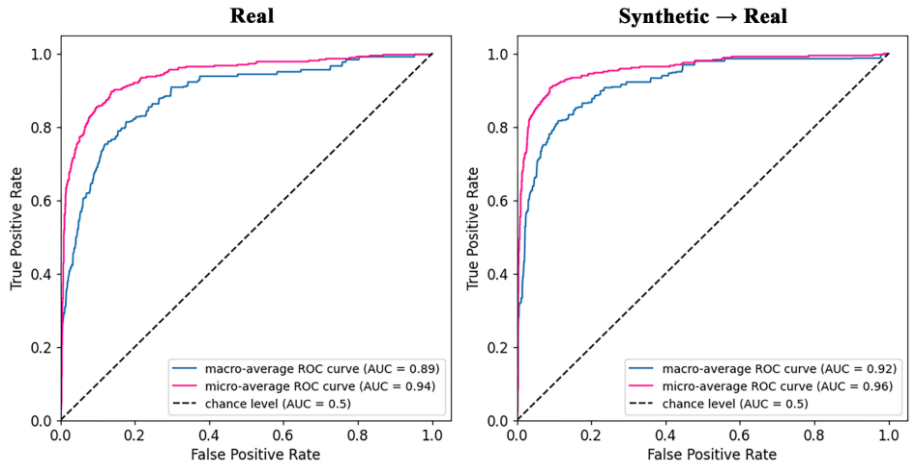

Figure 14: ROC curves for DR severity grading conducted across different experiments.

### 4.3.3. DR detection

In this section, comparative experiments are conducted to quantitatively evaluate the synthesized DR images specifically in the DR detection task. To set up the binary classification task, we first consider the images of the No-DR class as class 0, representing the healthy class, and then merge the images from all abnormal classes to create class 1, representing the unhealthy class. Firstly, similar to the experiments carried out for the grading task, we consider a baseline model trained only on real data (R). Subsequently, in another experiment (R+S (CStyleGAN)), we train the classifier on a combination of real and synthesized images generated by the conditional StyleGAN, where the size of the synthesized data is half that of the real data. Lastly, we adopt the novel effective SeFa-based data augmentation approach, wherein SeFa-based editing of synthesized images is utilized to effectively augment data, thereby further enhancing the classification performance. To achieve this, we collect an equal number of the synthesized abnormal images from different classes. By performing lesion manipulation using the SeFa algorithm, we try to obtain their healthy versions as shown in Fig.15. Subsequently, using the best multi-class classifier trained on real images, pseudo-labeling is conducted to ensure that we obtain zero-class images. The pairs consisting of the unhealthy image and its healthy manipulated version are added to the real images, resulting in the total number of fake images being half of the real images. Furthermore, in S → R experiments, the classifier is initially trained on synthesized images and then fine-tuned with real images. Table 3 shows the results of comparative experiments, and Table 4 demonstrates a comparison between our best result and some other state-of-the-art methods carried out for DR detection.

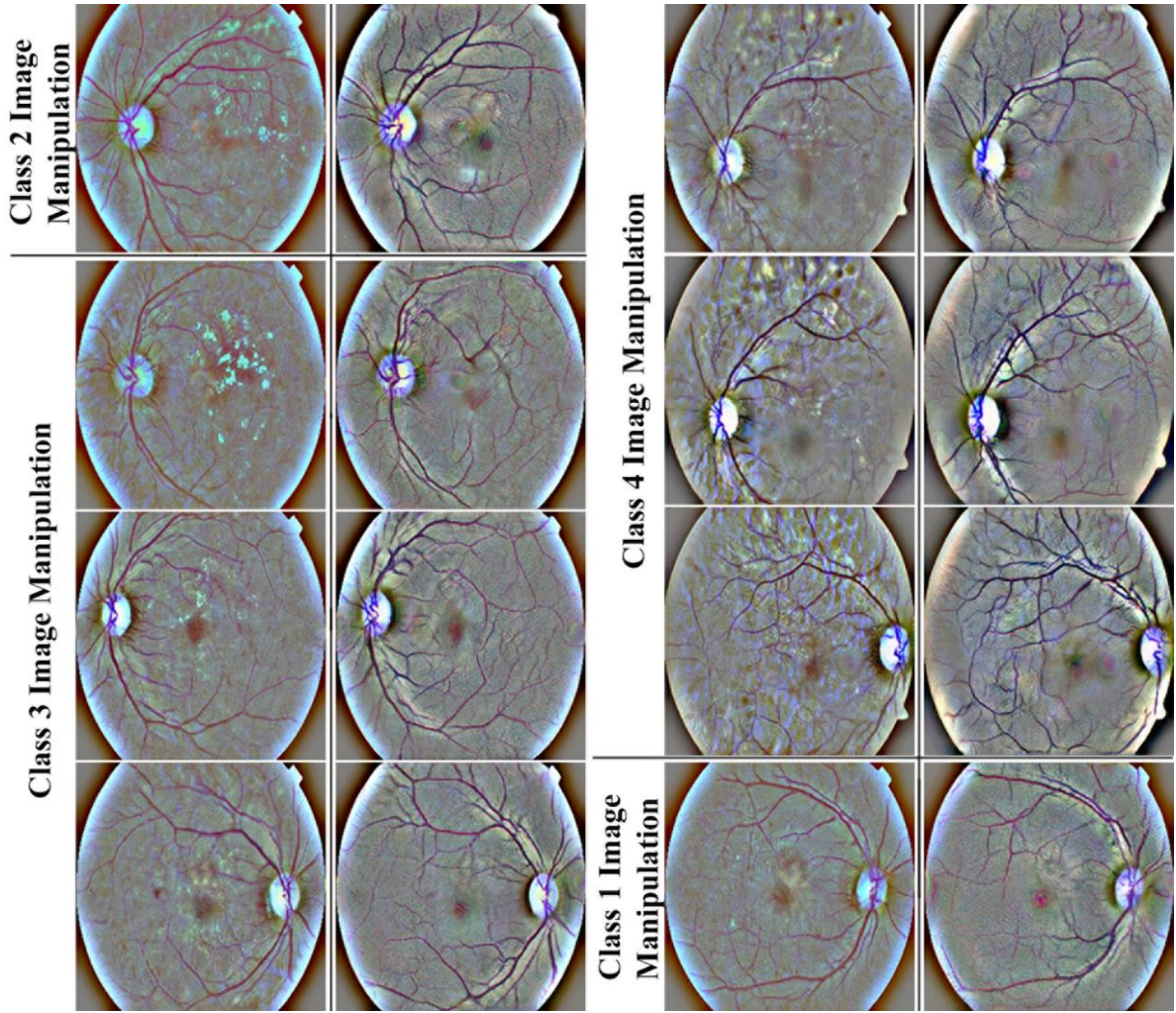

Figure 15: Samples synthesized using effective SeFa-based data augmentation. Each pair of images consists of the unhealthy retinal image on the left, synthesized by conditional StyleGAN, and the corresponding healthy version on the right, obtained through SeFa-based image manipulation.

Based on Table 3, incorporating images generated by conditional StyleGAN with existing real images enhances the classifier's performance in the DR detection task across most metrics. Remarkably, when the synthesized images stem from our proposed effective SeFa-based augmentation method, the metrics values experience an even more significant boost. Within this data augmentation framework, we manipulate lesions to modify solely the lesion areas of the image, leaving the rest unaltered. This enables the network to disregard redundant features and instead concentrate on crucial regions as discriminative features for DR identification.

Table 3: Performance comparison of the ResNet50 trained on different training sets for DR detection. The metrics are calculated based on macro average.

| Experiments | Accuracy | QWK | Sensitivity | Specificity | Precision | F1-score | AUC-ROC |
|---|---|---|---|---|---|---|---|
| R | 0.9754 | 0.9508 | 0.9624 | <u>0.9888</u> | <u>0.9890</u> | 0.9754 | 0.9756 |
| R + S(CStyleGAN) | <u>0.9781</u> | 0.9563 | <u>0.9731</u> | 0.9833 | 0.9837 | 0.9784 | 0.9782 |
| R + S(SeFa-based) | **0.9809** | **0.9618** | 0.9677 | **0.9944** | **0.9945** | <u>0.9809</u> | **0.9811** |
| S → R(SeFa-based) | **0.9809** | <u>0.9617</u> | **0.9785** | 0.9833 | 0.9838 | **0.9811** | <u>0.9809</u> |

In addition, Table 4 demonstrates that augmenting the existing real images with synthesized images from the effective SeFa-based data augmentation framework yields superior results compared to models employing more complex structures and approaches. Notably, the proposed framework exhibits significant advantages over previous methods utilizing attention structures [43], employing ensemble learning [45], or applying classical data augmentation [43] in terms of accuracy, specificity, precision, and F1-score. Fig.16 and 17 illustrate the confusion matrices and ROC curves, respectively, for DR detection across various experiments. It is evident that employing both real images and synthetic images synthesized through the effective SeFa-based augmentation technique for classifier training yields optimal outcomes. Specifically, this approach enhances the model's ability to identify DR disease with reduced misclassifications and increased discriminatory power between healthy (No DR) and unhealthy (DR) data.

Table 4: Comparison of the ResNet50 Performance with related methods for DR detection using APTOS 2019 dataset. The metrics are calculated based on macro average.

| Model | Accuracy | QWK | Sensitivity | Specificity | Precision | F1-score | AUC-ROC |
|---|---|---|---|---|---|---|---|
| Inception V3 [47] | 0.9359 | - | 0.93 | - | 0.9394 | 0.9347 | - |
| Am-Inception V3 [47] | 0.9446 | - | 0.90 | - | 0.989 | 0.9424 | - |
| CNN [48] | 0.946 | - | 0.86 | 0.96 | - | - | - |
| DenseNet169+CBAM+INS [43] | 0.97 | 0.9455 | **0.97** | 0.983 | - | - | - |
| ECOC-SVM [45] | 0.966 | - | 0.966 | 0.966 | 0.966 | 0.966 | - |
| RINet [49] | 0.95 | - | 0.95 | - | 0.95 | 0.95 | - |
| Ours | **0.9809** | **0.9618** | <u>0.9677</u> | **0.9944** | **0.9945** | **0.9809** | **0.9811** |

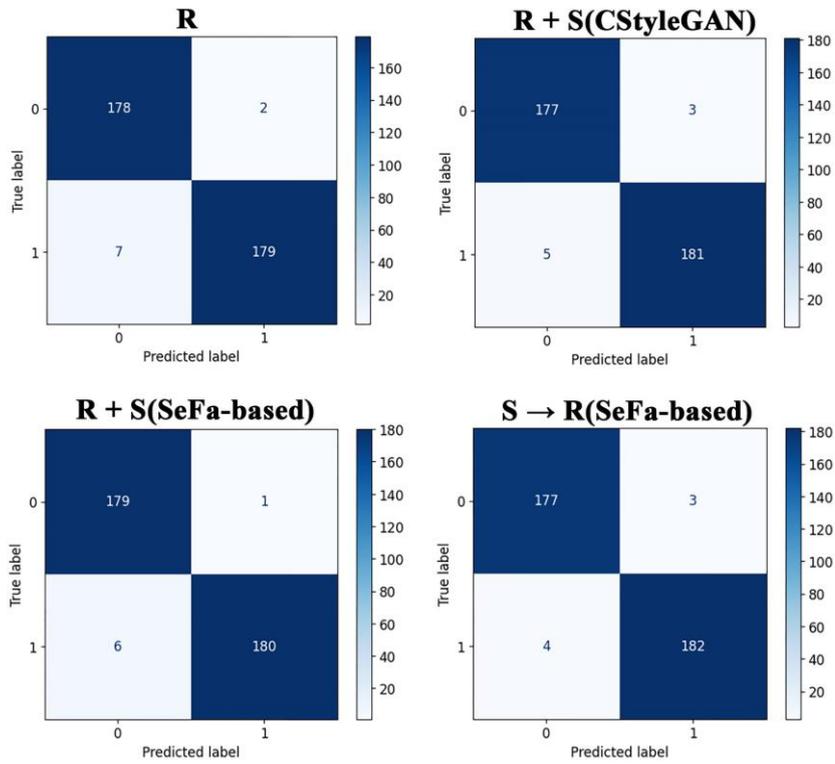

Figure 16: Confusion matrices for DR detection conducted across different experiments.

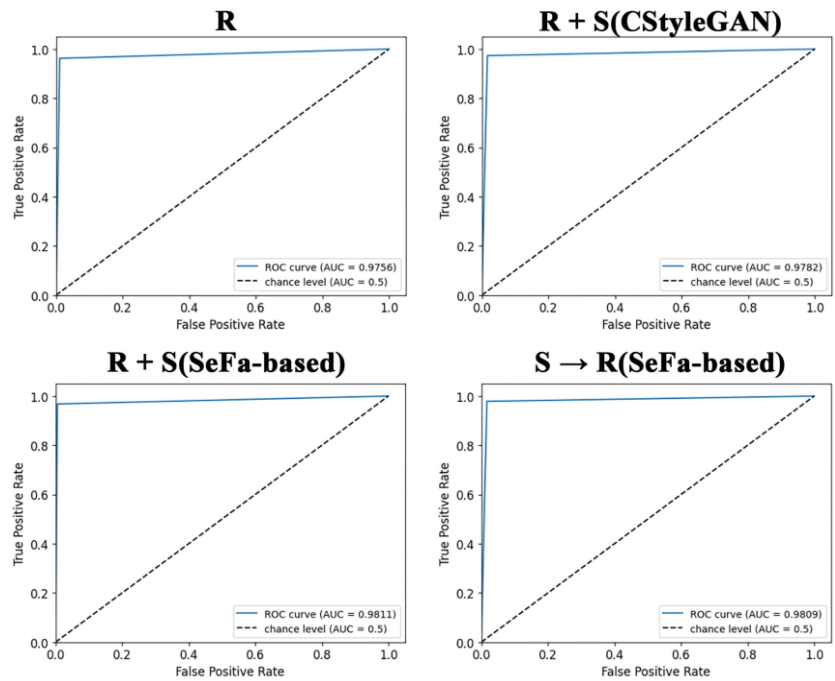

Figure 17: ROC curves for DR detection conducted across different experiments.

## 5. Discussion

This work proposes a comprehensive controller approach for synthesizing DR fundus images using a single GAN-based model. This approach aims to achieve meaningful data augmentation, leading to increased diversity of generated images and ultimately enhancing the performance of the classifier in both DR grading and detection tasks. Fig.5 showcases generated DR images by conditional StyleGAN, demonstrating initial control over DR grades resulting in highly realistic and high-quality synthetic images across different grades. Subsequent figures (Fig.6, Fig.7, Fig.8, and Fig.9) illustrate further control over the generated images from the previous step. This is achieved by manipulating various semantic features based on the significant directions discovered by the SeFa algorithm, including optic disc location, overall appearance, vessel structures, and lesion areas. Fig.10, depicting semantic feature manipulation based on Style mixing, confirms observations consistent with those of the SeFa algorithm, highlighting SeFa's superior control over such semantics. This demonstrates that integrating SeFa with the unique controller structure of StyleGAN, enhances control over synthetic DR images. The experimental results, presented in Tables 1 to 4, encompass a wide range of metrics, such as accuracy, QWK, sensitivity, specificity, precision, F1-score, and AUC. These results not only validate the benefits of incorporating synthesized images for training the classifier in both DR grading and detection tasks, but also demonstrate superiority over recent studies employing more complex architectures or training approaches. A key distinction of our controllable image generation approach from previous works [17], [18], [22] is that it eliminates the need for pre-existing feature masks or additional system training for mask generation. This simplifies the process and reduces complexity. Furthermore, by introducing a novel, effective SeFa-based data augmentation strategy, we provide a simple yet powerful tool for guiding the classifier to focus on discriminative regions while disregarding redundant features. Specifically, training the classifier solely with pairs of non-healthy images and their healthy versions synthesized through SeFa-based manipulation for DR detection yields promising results, as shown in Table 3.

## 6. Conclusion

The main aim of this study is to conduct comprehensively controllable DR fundus image generation, controlling various image aspects, including DR severity grades, optic disc appearance, vasculature patterns, and lesion characteristics. This fosters enhanced data diversity, ultimately leading to improved performance in DR grading and detection tasks. Recent studies have attempted to synthesize retinal images by controlling specific image features, requiring pre-existing masks for these features. This necessitates training an auxiliary network for mask generation, introducing complexity and dependence of the generated image's quality and diversity on the accuracy of these masks. To address these limitations, we utilized the controller structure of conditional StyleGAN and applied the SeFa method to discover meaningful directions within latent space for image manipulation. Additionally, we introduce a novel, effective SeFa-based data augmentation strategy. This strategy guides the classifier focus on discriminative image regions while disregarding redundant features. The qualitative and quantitative experimental results on the APTOS 2019 dataset showcase the generated images' ability to improve the performance of both DR grading and detection tasks and the model's superiority over other state-of-the-arts. One limitation of this study pertains to the generation of images with a low resolution of 256×256, thereby constraining both the quality of the generated images and the performance of the classifier. Consequently, for future research directions, there is a need to generate images with higher resolutions. This advancement would not only elevate the quality of the synthesized images but also enable more detailed manipulations. Specifically, manipulating vessel structures at higher resolutions would facilitate the analysis of diseases associated with retinal vessels, such as hypertensive retinopathy (HR), which could influence the progression of DR disease.

## Acknowledgments

This work received partial financial support from Amirkabir University of Technology (Tehran Polytechnic).

## CRediT authorship contribution statement

**Somayeh Pakdelmoez:** Conceptualization, Data curation, Investigation, Methodology, Software, Validation, Visualization, Writing – review and editing. **Saba Omidikia:** Conceptualization, Investigation, Methodology, Project administration, Validation, Writing – original draft, Writing – review and editing. **Seyyed Ali Seyyedsalehi:** Conceptualization, Funding acquisition, Supervision, Writing – review and editing. **Seyyede Zohreh Seyyedsalehi:** Supervision, Writing – review and editing.

## Declaration of generative AI and AI-assisted technologies in the writing process

During the preparation of this work the author(s) used ChatGPT in order to improve manuscript readability. After using this tool/service, the author(s) reviewed and edited the content as needed and take(s) full responsibility for the content of the published article.